# Solute-point defect interactions, coupled diffusion, and radiation induced segregation in fcc nickel

E. Toijer[1], L. Messina[2,1], C. Domain[3], J. Vidal[3], C.S. Becquart[4], and P. Olsson[1]

[1]*KTH Royal Institute of Technology, Nuclear Engineering, SE-114 21 Stockholm, Sweden*

[2]*CEA, DES, IRESNE, DEC, Cadarache F-13108 Saint-Paul-Lez-Durance, France*

[3]*Department of Materials and Mechanics of Components, EDF R&D, F-77250 Moret sur Loing, France*

[4]*Univ. Lille, CNRS, INRAE, Centrale Lille, UMR 8207 - UMET - Unité Matériaux et Transformations, F-59000 Lille, France*

## ABSTRACT

Radiation-induced segregation (RIS) of solutes in materials exposed to irradiation is a well-known problem. It affects the life-time of nuclear reactor core components by favouring radiation-induced degradation phenomena such as hardening and embrittlement. In this work, RIS tendencies in face-centered cubic (fcc) Ni-X (X = Cr, Fe, Ti, Mn, Si, P) dilute binary alloys are examined. The goal is to investigate the driving forces and kinetic mechanisms behind the experimentally observed segregation. By means of *ab initio* calculations, point-defect stabilities and interactions with solutes are determined, together with migration energies and attempt frequencies. Transport and diffusion coefficients are then calculated in a mean-field framework, to get a full picture of solute-defect kinetic coupling in the alloys. Results show that all solutes considered, with the exception of Cr, prefer vacancy-mediated over interstitial-mediated diffusion during both thermal and radiation-induced migration. Cr, on the other hand, preferentially migrates in a mixed-dumbbell configuration. P and Si are here shown to be enriched, and Fe and Mn to be depleted at sinks during irradiation of the material. Ti and Cr, on the other hand, display a crossover between enrichment at lower temperatures, and depletion in the higher temperature range. Results in this work are compared with previous studies in body-centered cubic (bcc) Fe, and discussed in the context of RIS in austenitic alloys.

## I. INTRODUCTION

Ni-based alloys and austenitic stainless steel are common structural materials in current and future generation nuclear power plants (NPPs). Novel material classes such as high-entropy alloys (HEA), or concentrated solid solution alloys (CSA) are also materials that can be Ni-based with a face-centered cubic (fcc) structure. Such materials can in the nuclear technology sector, be exposed to intense radiation fields, which can have a great impact on their properties. In Ni-based alloys, experimental observations show radiation-induced creep, swelling, and embrittlement [1], [2]. The change in material properties is generally related on the atomic scale to point-defect (PD) formation and diffusion. The coupling of PDs and atomic fluxes can lead to the redistribution of atoms independently of or alongside thermodynamic forces in a process known as radiation-induced segregation (RIS). This may in turn induce loss of strength and ductility, which consequently can lead to failure during reactor operation. Experimental characterization of irradiated material provides important insight on the processes involved in its degradation. However, the observations are in this case very much dependent on the exact conditions during irradiation. In order to better anticipate the evolution of an irradiated material, thermodynamic driving forces and solute-PD coupling must be understood, and atomic-scale modelling is a helpful tool for this task. The approach can provide important information on the interactions of various chemical species in metals, such as Ni-based alloys currently used in NPPs. Atomic-scale investigations of austenitic alloys and HEA, however, have been proven very difficult given their complex chemistry and magnetic properties. A common practice for modelling austenitic steels is to use Fe as a basis, since it represents the main constituent. However, replicating the fcc structure of austenitic alloys is difficult due to the instability of fcc Fe at 0-K, where Density Functional Theory (DFT) calculations are generally performed. This can be a problem since results obtained for one structure are not necessarily transferable to another [3]. Additionally, the paramagnetic state of fcc steels at higher temperatures is difficult to represent at 0-K where the ground state of Fe is body-centered cubic (bcc) and ferromagnetic. Proper modelling of austenitic alloys and HEAs remain for these reasons a challenging task.

In Ni-based, austenitic, and high entropy alloys, a number of different chemical species are present. For this reason, the impact of irradiation on the material cannot be understood only by looking at the behaviour of pure Ni. Multiple studies have been undertaken to clarify how common species behave in fcc Ni under irradiation. Fe and Cr have been shown to slow down defect migration and decrease the accumulation and average size of defect clusters in Ni alloys [4–10]. Si has been shown to suppress void formation and growth, and to enhance grain boundary cohesion [11–18]. P is known to embrittle austenitic alloys [19], but its embrittling effect on Ni is debated [17,18,20,21]. Studies regarding the behaviour of Ti in Ni alloys are scarcer; however, Ti depletion following irradiation has been observed [22], and solute-induced enhancement of theoretical strength has been reported [23]. Mn is present in all austenitic steels, as well as in Ni-based superalloys, and is also a common ingredient in fcc HEAs. For Mn in fcc Ni, RIS is indicated to be driven by vacancy exchange in the opposite direction to that of P [24]. From these previous works, it is clear that Fe, Cr, Si, P, Ti, and Mn can strongly impact the microstructural evolution of Ni alloys. For this reason, a correct description of RIS tendencies of each species is of high importance to anticipate the evolution of the irradiated system.

The solute RIS tendencies can be obtained by computing the full matrix of transport coefficients [25], which describe the kinetic response of each species to thermodynamic driving forces, and the possible presence of coupled fluxes. For fcc alloys, vacancy-solute transport coefficients can be obtained either by means of the Self-Consistent Mean-Field (SCMF) method [26,27] or the Green's-function method [28]. However, transport coefficients related to interstitial-assisted diffusion can be obtained only in a SCMF framework [29], which has been recently extended to arbitrarily long interaction ranges and any periodic crystal structure thanks to the KineCluE code [30]. Based on solute-PD binding and migration energies obtained with DFT, full dumbbell-transport matrices have been computed for the first time in bcc alloys [31], but never yet in fcc alloys. The approach extends the SCMF framework established for dumbbell diffusion in bcc and fcc alloys [29], and can be seen as a broad generalization of the traditional 5-frequency model [32], which is limited to vacancy diffusion and short-ranged solute-PD interactions. As such, it provides a complete framework by combining vacancy- and interstitial-assisted transport.

The goal of this study is to improve the current understanding of irradiated Ni alloys by an accurate description of RIS stemming from calculations of transport coefficients in dilute binary alloys. This has been done by applying the SCMF method parametrized with DFT calculations of solute-PD binding and migration energies to get a more accurate description of the behaviour of Fe, Cr, Si, Ti, Mn, and P solutes in fcc Ni. As the ground-state phase of Ni has an fcc structure, and austenitic steels used in today's NPPs feature a high Ni content (~10%), results obtained for fcc Ni are not only of interest for Ni-based alloys. Results here presented will, for this reason, be compared with previous results from bcc Fe and discussed in the context of radiation damage in austenitic steels.

# II. METODOLOGY

Atomic transport properties are examined in the framework of SCMF theory [26] using the KineCluE code [30]. The code allows for the calculation of transport coefficients, from which flux coupling and RIS tendencies can be determined, based on *ab initio* solute-defect thermodynamic interactions and migration barriers. In this section, the relevant background and methodology are outlined. Section 2.1 presents the theory of kinetic coupling and diffusion driving forces, section 2.2 gives a short description of how the transport properties were calculated in this work, and in section 2.3, the *ab initio* methodology is presented.

### A. Kinetic Coupling and transport coefficients

Even though vacancies occur naturally in any metallic material, in a reactor at its operating temperatures (~300º C for Gen II-III, ~600º C for Gen IV), the vacancy population is dominated by irradiation induced effects. In addition, the radiation field is responsible for the generation of an equal amount of self-interstitial atoms (SIAs) that are essentially negligible under thermal conditions. The spatially inhomogeneous formation of defects results in chemical potential gradients (CPG), which in turn induce atomic and defect fluxes in the material. The flux of a species α due to the CPGs acting on each species β is given by Eq. 1.

$$J_\alpha = -\sum_\beta L_{\alpha\beta} \frac{\nabla \mu_\beta}{k_B T} \qquad (1)$$

where $L_{\alpha\beta}$ are the transport coefficients (also known as the coefficients of the Onsager matrix), i.e., the proportionality factors between the flux of a species α and a CPG, $\nabla\mu$, acting on species β. In a binary alloy, α = *A, B, V*, and *I*, which represent solvent atoms (Ni), solutes, vacancies, and interstitial defects respectively. In this case, there are three independent coefficients for each defect, namely $L_{VV}$, $L_{VB}$, $L_{BB}(V)$ for vacancy-assisted diffusion, and $L_{II}$, $L_{IB}$, $L_{BB}(I)$ for interstitial-assisted diffusion. These coefficients describe the kinetic response of a system to thermodynamic forces. The off-diagonal coefficients, $L_{VB}$ and $L_{IB}$, describe the coupling between different species, i.e., when a flux of a given species is induced by the CPG acting on another species. In addition, the knowledge of the full Onsager matrix allows for the calculation of tracer self- and solute diffusion coefficients, as well as for the prediction of solute radiation-induced segregation tendencies.

*1. Radiation-induced segregation*

If the transport coefficients are known, it is possible to model RIS under steady-state conditions in a binary alloy (*AB*) using Eq. 2 [31,33,34].

$$\frac{\nabla c_B}{c_B} = -\alpha \frac{\nabla c_V}{c_V} \qquad (2)$$

where

$$\alpha = \frac{L_{IA} L_{VA}}{\phi(L_{IA} D_B + L_{IB} D_A)} \left( \frac{L_{VB}}{L_{VA}} - \frac{L_{IB}}{L_{IA}} \right) \qquad (3)$$

where $D_A$ and $D_B$ are the intrinsic diffusion coefficients of solvent and solute atoms. These can be computed directly from the transport coefficients (the full set of equations are provided in a previous publication [31]). The thermodynamic factor, $\phi$, describing the change of chemical potential of one species with respect to a concentration change of another, can be assumed equal to unity in the dilute limit. The model in Eqs. 2-3 represents the steady-state solution of Eq. 1, and assumes a low defect-sink density and the absence of relevant sink bias. In addition, accordingly to the dilute limit, multiple-solute and multiple-defect effects are neglected. More details about the derivation and the underlying assumptions of the model can be found in previous publications [31,33,34]. From Eq. 2 it can be seen that the sign of α determines if the concentration gradient of species *B* is in the same direction as that of defects. Based on the assumption that defects diffuse towards sinks, where they are preferentially absorbed, the corresponding concentration gradient is negative, so a positive α indicates enrichment of species *B* at sinks, whereas a negative α indicates depletion. In Eq. 3, the sign-determining factor of α is given by

$$\left( \frac{L_{VB}}{L_{VA}} - \frac{L_{IB}}{L_{IA}} \right)$$

where the left fraction is related to vacancy-coupled fluxes, and the right one to SIA-coupled ones. From the two fractions, it is possible to investigate the segregation tendencies induced by each mechanism independently. The two fractions are termed Partial Diffusion Coefficient (PDC) ratios, and the impact of the respective mechanisms is determined by setting the opposite PDC ratio to 1, describing in this case an uncorrelated flux between the solute and that defect. In the case of vacancy-mediated diffusion, the competition between the solute and the bulk species can result in three distinct cases. In the case of preferential vacancy-solute exchange and positive flux coupling ($L_{BV} > 0$), solutes migrate in the same direction as vacancies, and solute enrichment at sinks occurs. This process is known as vacancy drag, and is indicated by a negative vacancy PDC ratio since $L_{AV}$ is always negative. When vacancy drag does not take place, enrichment can still take place, as in this case, both solvent and solute atoms diffuse against the vacancy flux (inverse Kirkendall mechanism), thus away from the sink. If the solute is slower than the solvent (preferential solvent-vacancy exchange), the solute will effectively be enriched at sinks. This is indicated by $0<PDC_{vac}<1$. In case, instead, of preferential solute-vacancy exchange, solutes will diffuse away faster than solvent atoms, and depletion occurs ($PDC_{vac} > 1$). For interstitial-mediated diffusion, the flux of solutes cannot be in the opposite direction to that of interstitials, so the $PDC_{SIA}$ is always positive. If solute transport is faster than solvent transport, this results in enrichment at sinks, and in this case the $PDC_{SIA}$ is greater than 1. The $PDC_{SIA}$ is smaller than 1 in the opposite case.

## 2. Self- and solute diffusion coefficients

Solute and solvent diffusion coefficients in thermodynamic equilibrium have been determined experimentally for many systems. As diffusion coefficients can also be calculated based on the transport coefficients obtained in the context of SCMF theory, they are useful for validation of results. Under thermal-equilibrium conditions, migration occurs predominantly by a vacancy-assisted mechanism, and the self (solvent) and solute diffusion coefficients can be computed with Eq. 4 and 5 respectively [32].

$$D^* = g a_0^2 f_0 \omega_0 c_v^{eq} = g a_0^2 f_0 \nu_0 \exp\left(-\frac{E_v^f + E_v^m}{k_B T}\right) \exp\left(\frac{S_v^f}{k_B}\right) \qquad (4)$$

$$D_B^* = g a_0^2 f_B \omega_2 c_V^{eq} p_{1nn} = g a_0^2 f_B \nu_B \exp\left(\frac{-E_V^f - E_B^m + E_{BV}^{b,1nn}}{k_B T}\right) \exp\left(\frac{S_V^f}{k_B}\right) \qquad (5)$$

where $g$ is a geometrical factor (g=1 for monovacancy diffusion in fcc), $a_0$ is the lattice constant, $f_0$ and $f_B$ are the self- and solute correlation factors, $\omega_{0,B}$ are the jump frequencies of a vacancy exchanging respectively with a solvent atom (in the absence of solutes nearby) or a solute atom, $v_{0,B}$ are the corresponding attempt frequencies, $p_{1,nn}$ is the probability of having a solute-vacancy pair at a first nearest neighbour (1nn) distance, $E_V^f$ and $S_V^f$ are the vacancy formation energy and entropy, $E_B^m$ is the solute migration barrier, $E_{BV}^{b,1nn}$ is the 1nn solute-vacancy binding energy, here defined as attractive when positive. The solute correlation factor, $f_B$, is related to the probability for the solute atom to make an immediate reverse jump back to its previous position, thus leading to no net displacement. In the context of the 5-frequency model [32], this factor is obtained by considering only the probabilities of the defect returning from second, third and fourth nearest neighbouring positions to the 1nn position with respect to the solute [32]. In place of Eqs. 5 and 6, the diffusion coefficients can be computed directly from the transport coefficients according to Eq. 7 and 8 (valid in the case of dilute concentrations), in which $A^*$ is the solvent tracer.

$$D^* = \frac{L_{A^*A^*}}{c_{A^*}} \tag{6}$$

$$D_B^* = \frac{L_{BB}}{c_B} \tag{7}$$

The resulting correlation factors $f_0$ and $f_B$ are included in $L_{A^*A^*}$ and $L_{BB}$ respectively. Within SCMF theory and its implementation in the KineCluE code, the calculation of the transport coefficients considers kinetic trajectories of increasing amplitude, up to a cut-off kinetic radius that can be arbitrarily chosen by the user. In addition, whereas in Eq. 5 the probability of a vacancy-solute pair accounts for 1nn thermodynamic interactions only, KineCluE allows for a more accurate evaluation of the pair partition function, thus providing a pair probability that takes into account longer-distance vacancy-solute interactions. Thanks to this approach, it is thus possible to provide a more accurate evaluation of the kinetic properties of Ni alloys with respect to previous works based on the 5-frequency model.

## B. Calculation of transport properties

The symmetric Onsager matrix is calculated in this work using the KineCluE code, which implements SCMF theory to expand the Onsager matrix in terms of cluster contributions [35]. The transport coefficients are in this case given by Eq. 8.

$$L_{ij} = \zeta \sum_c f_c L_{ij}^{(c)} \tag{8}$$

where $\zeta$ is the total concentration of all monomers and clusters and $f_c$ is the concentration fraction of cluster $c$. For a full discussion of the breakdown of transport coefficients into cluster contributions, see [31] and [35]. In the dilute limit, the Onsager matrix is split into contributions from isolated vacancies (*V*), isolated interstitials (*I*), solute-vacancy pairs (*VB*), and solute-interstitial pairs (*IB*). The matrix is in this case given by Eq. 9 [31].

$$\begin{bmatrix} L_{VV} & L_{VI} & L_{VB} \\ L_{IV} & L_{II} & L_{IB} \\ L_{VB} & L_{IB} & L_{BB} \end{bmatrix} = \zeta \left( f_V \begin{bmatrix} L_{VV}^{(V)} & 0 & 0 \\ 0 & 0 & 0 \\ 0 & 0 & 0 \end{bmatrix} + f_I \begin{bmatrix} 0 & 0 & 0 \\ 0 & L_{II}^{(I)} & 0 \\ 0 & 0 & 0 \end{bmatrix} + f_{VB} \begin{bmatrix} L_{VV}^{(VB)} & 0 & L_{VB}^{(VB)} \\ 0 & 0 & 0 \\ L_{VB}^{(VB)} & 0 & L_{BB}^{(VB)} \end{bmatrix} + f_{IB} \begin{bmatrix} 0 & 0 & 0 \\ 0 & L_{II}^{(IB)} & L_{IB}^{(IB)} \\ 0 & L_{IB}^{(IB)} & L_{BB}^{(IB)} \end{bmatrix} \right) \quad (9)$$

where $f_X$ represent the corresponding cluster concentrations. Interstitial-vacancy correlations represented by the $L_{IV}=L_{VI}$ coefficients are neglected in this work, under the assumption that the recombination probability is low when the concentrations of the two types of point defects are sufficiently low.

In order to calculate the transport coefficients of Eq. 9, thermodynamic interactions and migration mechanisms must be outlined. In the case of interstitial migration, three possible configurations should be considered in the fcc lattice; octahedral, tetrahedral, and dumbbell. DFT calculations in this work demonstrate that, the ⟨100⟩ dumbbell is associated in Ni with the lowest energy of the possible interstitial configurations for all species considered, with the notable exception of P. Indeed, it is shown in section 3.1 that the mixed-P ⟨100⟩ dumbbell displays a severe instability, and that the octahedral configuration is significantly more stable for that species. DFT calculations in this work also demonstrate that if a pure Ni-dumbbell comes sufficiently close to a substitutional P, the P will be kicked out into a octahedral configuration, as one of the Ni atoms of the dumbbell takes its place in the lattice. For this reason, only dumbbell-induced octahedral diffusion is relevant in the case of P interstitials. As the P atoms need the presence of a pure Ni-dumbbell to be kicked out into an octahedral configuration, the diffusion process is strongly coupled with pure-Ni dumbbell diffusion. Regarding all other solutes in this work, DFT results indicate that the ⟨100⟩ dumbbell is of main importance for interstitial migration. In this configuration, a solute can be part of the defect as a mixed dumbbell, or in its neighbourhood. Concerning 1nn configurations, which are characterized by the strongest thermodynamic interactions, the solute can be in a compressed (a-type) or tensed (b-type) position depending of the dumbbell orientation in relation to the solute. The dumbbell can for this reason move in multiple ways in relation to the solute. The ⟨100⟩ mixed-dumbbell and 1nn configurations considered in this work are illustrated in Fig 1. As it is reasonable to assume that the solute-dumbbell interaction quickly drops to zero after the 1nn, as was the case for Fe alloys [31], interactions beyond this distance were not explicitly calculated in this work.

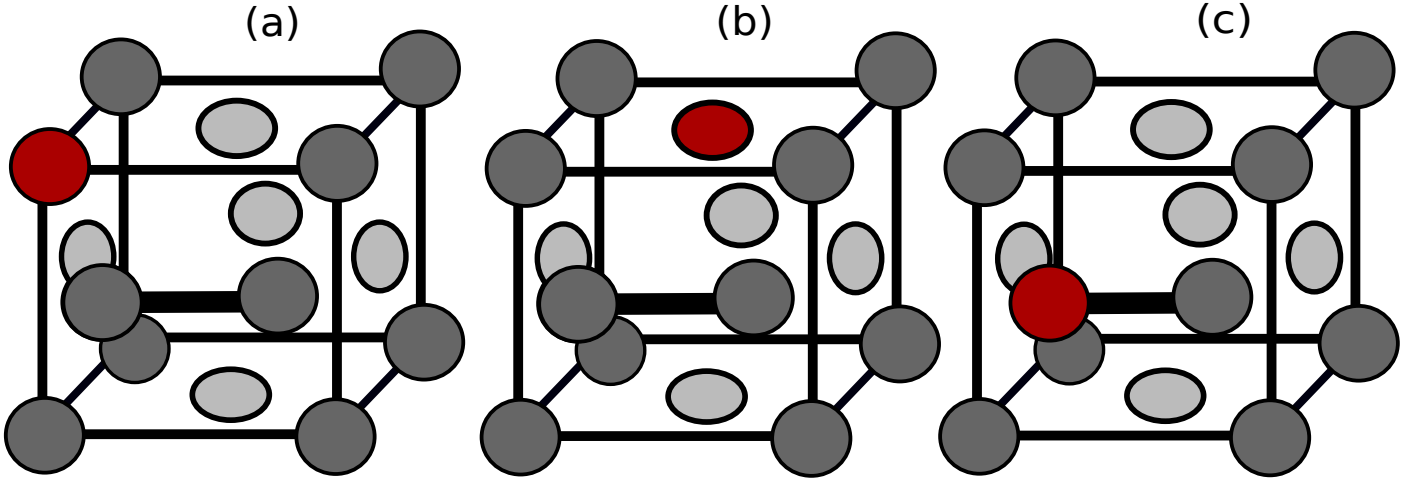

FIG. 1. Three ⟨100⟩ dumbbell configurations in an fcc lattice. a) A pure nickel dumbbell with the solute in a compressed lattice site (a-type) b) A pure nickel dumbbell with the solute in a tensed lattice site (b-type). c) A mixed dumbbell. The red circles indicate the position of the solute atom.

After the thermodynamics of the system has been analysed, the corresponding migration paths of the energetically favourable configurations are determined in order to calculate the transport coefficients. The set of symmetry-unique configurations and jump-frequencies needed for the calculation of the transport coefficients can be determined either explicitly using *ab initio* (DFT), or by the Kinetically-Resolved Activation (KRA) approach [36] implemented in the KineCluE code. The jumps which are explicitly calculated with DFT in this work are illustrated in Fig. 2 in the case of solute-vacancy related migration, and Fig. 3 in the case of solute-⟨100⟩ dumbbell related migration. KineCluE evaluates all other possible jumps up to a maximum trajectory range, i.e. the kinetic radius, via the KRA approach. As the transport coefficients have been shown to be well converged within a kinetic radius of $4a_0$ [30,31], this value was used in all calculations.

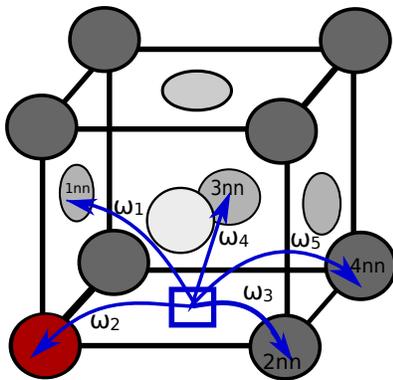

FIG. 2. Illustration of the solute-vacancy related migration barriers explicitly calculated using DFT. The red circle indicates the position of the solute atom, the blue square that of the vacancy. The neighbour ordering is with respect to the solute atom.

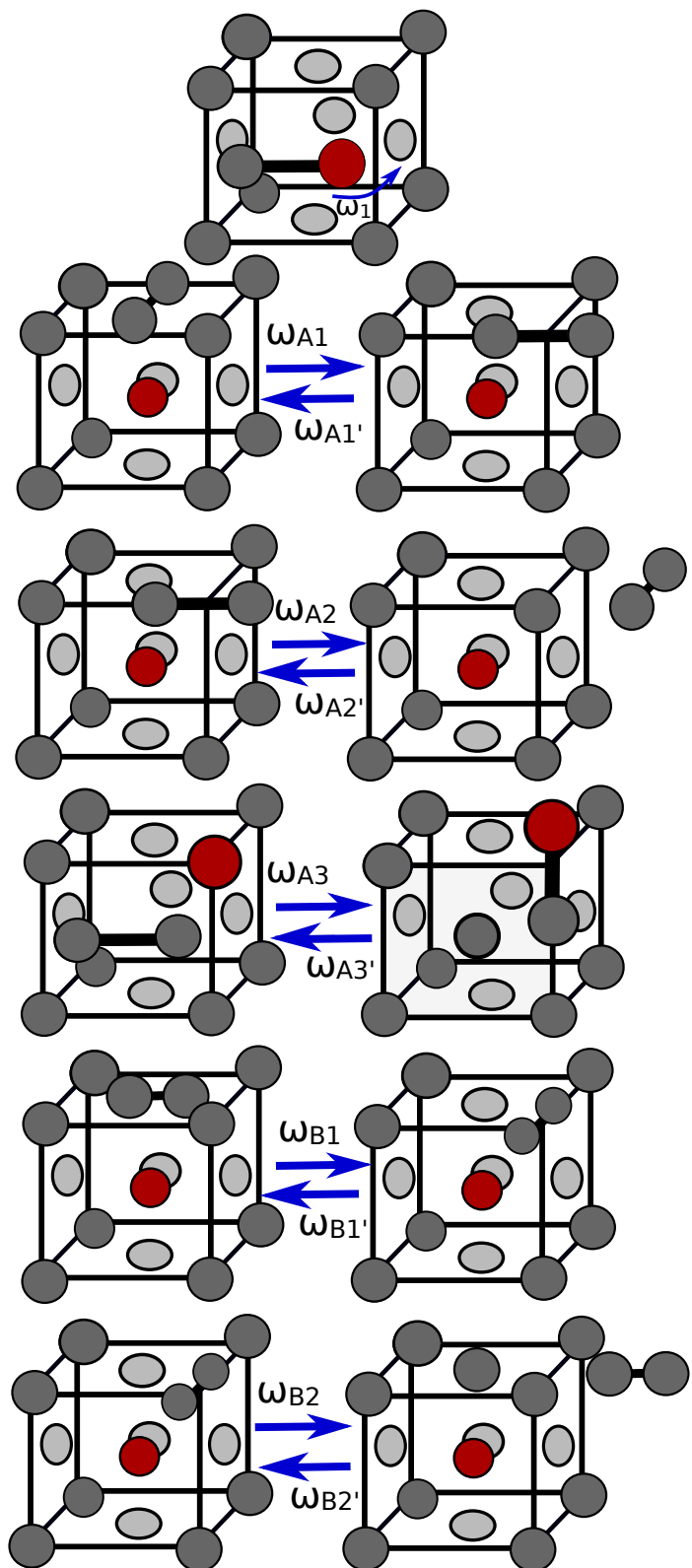

FIG. 3. Illustration of the solute-dumbbell related migration barriers explicitly calculated using DFT. The red circle indicates the position of the solute atom.

Once the thermodynamic interactions and migration mechanisms have been evaluated, the cluster transport coefficients of Eq. 9 can be calculated with KineCluE, following the process outlined in [31]. Calculations were performed in the dilute limit, with a solute-to-solvent ratio of $10^{-4}$. As the transport coefficients are given as output from KineCluE calculations, RIS tendencies of the system are thereafter evaluated following the approach of section 2.1.

## C. *Ab initio* methodology

The calculation of transport coefficients relies on thermodynamic parameters (formation and binding energies), and kinetic parameters (jump frequencies).

For a configuration containing $n$ atoms of the solvent $X$ and $p$ sites of the solute $Y$, the formation energy is given by Eq. 10.

$$E_f(nX+pY) = E[nX+pY] - nE[X] - pE_{imp}[Y] \qquad (10)$$

Where $E_{imp}$ is the reference state of the solute. The crystal structures of the used reference states are: body centered cubic (bcc) for Cr, Mn, and Fe, tetrahedral structure (p4) for P, diamond structure for Si and hexagonal closed packed (hcp) structure for Ti.

The binding energy between $n$ objects in a supercell is given in Eq. 11.

$$E_b(A_1, A_2, ..., A_n) = \sum_{i=1,...,n} E(A_i) - \left(E_{ref} + E(A_1+A_2+...+A_n)\right) \qquad (11)$$

where $E_{ref}$ is the energy of the supercell without any defects, $E(A_i)$ is the energy of the supercell with the isolated defect $A_i$, and $E(A_1+A_2+...+A_n)$ is the energy of the cell containing all $A_i$ interacting defects. With this definition, positive values are binding (or attractive) configurations.

The jump frequency is defined by Eq. 12.

$$\omega = \nu \exp(-E_{mig}/k_B T) \qquad (12)$$

where $E_{mig}$ is the migration barrier and $\nu$ the attempt frequency, given by Eq. 13.

$$\nu = \frac{\prod_{j=1}^{3N-3} \nu_j^R}{\prod_{j=1}^{3N-4} \nu_k^S} \tag{13}$$

where $\nu_j^R$ are the vibrational eigenfrequencies in the relaxed defect supercell, and $\nu_j^S$ the eigenfrequencies in the saddle-point configuration.

The above properties (formation, binding and migration energies, and attempt frequencies) were obtained through *ab initio* density functional theory (DFT) calculations using the Vienna Ab initio Simulation Package (VASP) [37,38], with pseudopotentials from the VASP library generated with the projector augmented wave (PAW) method using the Perdew-Burke-Ernzerhof (PBE) exchange-correlation functional [39]. Spin polarization, periodic boundary conditions, and the supercell approach were applied for all calculations. The Brillouin zone was sampled with 3×3×3 k-points using the Monkhorst-Pack scheme. Supercells of 256 (4×4×4) fcc lattice sites were used for the calculations, unless otherwise stated. The plane-wave energies were cut off at 350 eV, and all relaxations were performed under constant volume conditions with a Ni lattice parameter of 3.522 Å. Calculations were performed in the ferromagnetic state; however, as magnetic moments of Cr solutes can be sensitive to the initial state, an initial guess of $-2\mu_B$ was used for this species to obtain the lower energy state. In order to calculate migration barriers, the Nudged Elastic Band method (NEB) [40,41] implemented in VASP was used with three images and the climbing-image algorithm [42] to obtain the saddle-point energy. This was checked to be sufficient for a precision of 20 meV/Å in force convergence. Attempt frequencies for vacancy migration were determined by the finite-difference method using the Phonopy software [43] in 256(±1) atom supercells with a displacement per atom of ±0.01Å. The initial structure and the structure at the saddle point were relaxed with a high accuracy in terms of residual forces ($10^{-7}$ eV on the total energy, and each force below 0.3 meV/Å).

# III. RESULTS

## A. Point Defect Equilibrium Properties

The defects considered in this work are the mixed and pure ⟨100⟩, ⟨110⟩, and ⟨111⟩ dumbbells, the octahedral, tetrahedral, and substitutional impurities, as well as the vacancy in pure Ni. The defect formation energies in bulk Ni are presented in Table I, and the vacancy-solute and self interstitial atom (SIA)-solute binding energies are presented in Table II. Vacancy-solute binding energies are presented as functions of the mutual distance between the two. Solute interactions with the pure ⟨100⟩ Ni dumbbell were also considered with a solute atom as a first nearest-neighbour (1nn) in compressed and tensed lattice sites. For an illustration of the considered solute-⟨100⟩ dumbbell configurations, see Fig. 1.

TABLE I: Point-defect and impurity formation energies (eV) in Ni. Unstable configurations are omitted and marked with "-". See Fig. 1 for an illustration of the three ⟨100⟩ dumbbell configurations. All calculations were performed in a 108-atom cell and with an energy cut-off of 350 eV.

| Defect | Ni | Cr | Fe | P | Si | Ti | Mn |
|---|---|---|---|---|---|---|---|
| ⟨100⟩ | 4.2 | 3.9 | 3.8 | - | 1.8 | 3.0 | 4.0 |
| ⟨110⟩ | 5.0 | 4.6 | 4.8 | 1.1 | 2.6 | 3.4 | 4.6 |
| ⟨111⟩ | 4.8 | 4.2 | 4.4 | 1.4 | 2.6 | - | - |
| Octahedral | 4.4 | 4.1 | 4.0 | 0.2 | 1.9 | 3.03 | 4.4 |
| Tetrahedral | 4.8 | 4.2 | 4.3 | 1.5 | 2.9 | 2.96 | 4.4 |
| Substitutional | - | 0.39 | -0.40 | -1.8 | -1.6 | -1.4 | -0.5 |
| Vacancy | 1.4 | - | - | - | - | - | - |
| A-type ⟨100⟩ (Compressed) | - | 4.2 | 3.8 | - | 2.2 | 2.75 | 3.7 |
| B-type ⟨100⟩ (Tensed) | - | 4.3 | 3.8 | 2.9 | 3.0 | 2.45 | 3.5 |

TABLE II: Solute-defect binding energies (eV) in pure Ni; positive values represent binding configurations. Unstable configurations are omitted and marked with "-". See Fig. 1 for an illustration of the three ⟨100⟩ dumbbell configurations. The binding energy for vacancy-solute binding is given as a function of nearest neighbour (nn) distance between the two. In the case of the octahedral and tetrahedral configurations, the ⟨100⟩ pure dumbbell and the substitutional solutes were used as reference.

| | Cr | Fe | P | Si | Ti | Mn |
|---|---|---|---|---|---|---|
| 1nn | -0.04 | -0.02 | 0.3 | 0.1 | 0.05 | -0.01 |
| 2nn | 0.02 | -0.01 | 0.03 | 0.0 | -0.06 | -0.03 |
| 3nn | -0.03 | -0.03 | -0.03 | -0.03 | -0.03 | -0.02 |
| 4nn | 0.02 | 0.03 | 0.01 | 0.01 | 0.1 | 0.03 |
| 5 nn | 0.0 | -0.01 | -0.01 | -0.01 | -0.01 | -0.01 |
| 6 nn | 0.0 | 0.0 | 0.0 | 0.0 | 0.0 | 0.0 |
| Mixed ⟨100⟩ | 0.4 | -0.1 | - | 0.8 | -0.3 | -0.3 |
| **Mixed ⟨110⟩** | -0.3 | -0.8 | 1.2 | -0.04 | -0.7 | -1.0 |
| **Mixed ⟨111⟩** | 0.1 | -0.6 | 1.0 | 0.01 | - | - |
| **A-type ⟨100⟩ (Compressed)** | 0.1 | -0.02 | - | 0.4 | -0.1 | -0.1 |
| **B-type ⟨100⟩ (Tensed)** | 0.0 | 0.02 | -0.5 | -0.4 | 0.02 | 0.2 |
| **Octahedral** | 0.2 | -0.3 | 2.1 | 0.7 | -0.3 | -0.5 |
| **Tetrahedral** | 0.1 | -0.5 | 0.9 | -0.2 | -0.3 | -0.6 |

In this work the vacancy formation energy in Ni was calculated to 1.4 eV. This value is in line with calculations by Nazarov *et. al* where the PBE exchange correlation functional was also used [44]. Results in Table I show that the preferred SIA configuration in pure Ni is the ⟨100⟩ dumbbell. This is in contrast to bcc Fe, where the ⟨110⟩ dumbbell is the most stable [31,45,46]. The introduction of Cr and Si solute atoms in Ni does not change the relative stability of the ⟨100⟩ dumbbell. From the strong stability of the mixed ⟨100⟩ dumbbells for the two species, one can suspect an efficient solute transport due to this migration mechanism. The strong interaction of the mixed Si-Ni dumbbell is different from observations of Si in bcc Fe, where the Fe-Si dumbbell is neither binding nor repulsive [31]. For Cr, on the other hand, the formation of a stable ⟨100⟩ dumbbell complex is in line with observations in bcc Fe, where the species has been shown to form a stable mixed ⟨110⟩ dumbbell [31]. Additionally, Cr has been shown to form a stable mixed ⟨100⟩ dumbbell in fcc Fe (antiferromagnetic matrix) [47] and in fcc FeNiCr (special quasirandom structure) [48]. This indicates that the interstitial migration of Cr may show a similar character in Fe and Ni. On the other hand, the introduction of Fe, P, Ti, and Mn impacts the relative stability of the ⟨100⟩ dumbbell. As shown in Table II, Fe, Ti, and Mn display repulsive interstitial interactions in all but the B-type 1nn configuration, where the binding is anyway very weak. Note that these results apply to 0-K ground-state properties only, and relative stabilities may be altered by finite-temperature effects. In the case of Ti and Mn, however, the mixed ⟨100⟩ dumbbells are particularly repulsive. As a consequence, the dissociation of the mixed dumbbell is far more likely than its migration, and a net solute displacement induced by the interstitial mechanism is unlikely. For this reason, interstitial migration of Ti and Mn will not be considered in this work. The result is in contrast to what has been observed in bcc Fe, where Mn has been shown to form stable mixed dumbbells and migrate efficiently via the dumbbell mechanism [31]. Regarding P, it can be seen in Tables I and II that the species displays the highest stability and strongest binding as an octahedral compared to any other interstitial configuration. This indicates that the mixed-dumbbell mechanism is likely not dominant in the migration of P atoms. Indeed, DFT calculations in this work evidenced that if the pure Ni dumbbell comes sufficiently close to a P atom in a substitutional site, the P will be subjected to a 'kickout' mechanism, where the P ends up in an octahedral configuration, while being replaced in the original substitutional site by one of the Ni atoms of the dumbbell. As a consequence, the main interstitial migration path of P will be through octahedral sites, once the kickout has taken place. In bcc Fe, P has been proven most stable in a mixed ⟨110⟩ dumbbell configuration [31,49,50]. Although migration of the ⟨110⟩ mixed dumbbell in Fe takes place in competition with octahedral migration [51], the lack of stability of the mixed Ni-P dumbbell evidenced in this work indicates a very different behaviour of P in the two materials.

The vacancy-solute binding energies from Table II are illustrated in Fig. 4. It can be seen that P displays the strongest binding, followed by Si and Ti. For this reason, one can suspect vacancy drag of the three species. It should however be noted that the interaction ranges for all vacancy-solute pairs lie in a very short span, and can be considered negligible beyond the 2nn. Cr, Fe, and Mn all display a weak negative coupling with vacancies, indicating that vacancy drag is less likely for the three species. In bcc Fe; Mn, P, Si, Cr, and Ni have been shown to form stable vacancy-solute pairs [31,49]. The weak repulsion between the vacancy and Cr is however in agreement with results from DFT calculations in fcc Fe [47] and fcc FeNiCr [48]. Overall, the solute-defect interactions in fcc Ni are different from those in bcc Fe, with the exception of the P/Si-vacancy interactions, the mixed Cr-Ni/Fe-dumbbells, and the Ni-Fe dumbbells, which are stable in both materials.

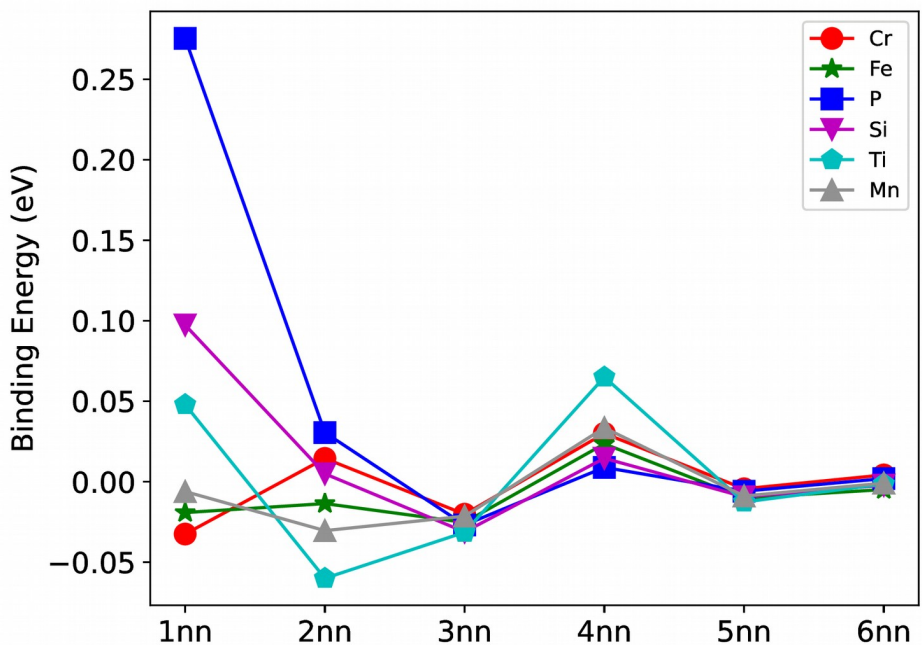

FIG. 4. Vacancy-solute binding energies (eV) in pure Ni as a function of nearest neighbour (nn) distance. Positive values represent binding configurations.

## B. Point Defect Migration Kinetics

The migration barriers for vacancy-mediated diffusion of solutes in fcc Ni are presented in Table

III.

TABLE III: Vacancy migration energies ω [eV] in the vicinity of various solutes in Ni, and solute-jump attempt frequencies, ν [THz]. See Fig. 2 for an illustration of the jumps and description of notations. Calculations marked with "*" were performed in a 108-atom cell, all others were performed in 256-atom cell. An energy cut-off of 350 eV was used in all calculations.

|  | Cr | Fe | P | Si | Ti | Mn |
|---|---|---|---|---|---|---|
| $\nu_0$ for vacancy jump $\omega_0$ | 14.32 | | | | | |
| $\omega_0$ Ni Self migration | 1.05 | | | | | |
| $\nu_2$ for solute jump $\omega_2$ | 10.85 | 12.45 | 0.67 | 2.64 | 7.07 | 6.77* |
| $\omega_2$ solute jump | 0.77 | 0.93 | 0.58 | 0.79 | 0.50 | 0.75 |
| $\omega_1$ 1nn → 1nn | 1.04 | 1.09 | 0.70 | 0.91 | 1.26 | 1.19 |
| $\omega_3$ 1nn → 2nn | 1.09 | 1.05 | 1.28 | 1.17 | 1.02 | 1.02 |
| $\omega_3'$ 2nn → 1nn | 1.14 | 1.05 | 1.03 | 1.07 | 0.91 | 1.00 |
| $\omega_4$ 1nn → 3nn | 1.12 | 1.03 | 1.13 | 1.10 | 1.00 | 1.02 |
| $\omega_4'$ | 1.13 | 1.03 | 0.83 | 0.98 | 0.92 | 1.01 |

| | Cr | Fe | P | Si | Ti | Mn |
|---|---|---|---|---|---|---|
| 3nn → 1nn | | | | | | |
| $\omega_5$ 1nn → 4nn | 1.11 | 1.03 | 1.11 | 1.08 | 0.98 | 1.01 |
| $\omega_5'$ 4nn → 1nn | 1.17 | 1.07 | 0.84 | 1.00 | 1.00 | 1.05 |

Results in Table III show that the attempt frequency for the vacancy-Fe jump is similar to that of pure Ni; it is decreasingly lower for Cr, followed by Ti, Mn, and Si. P has a considerably lower attempt frequency compared to the other solutes. The observations are in line with the Meyer-Nelder rule, according to which a high attempt frequency tends to compensate for higher barriers [52]. Ti and Si however, seem to contradict this rule. The trends observed in Table III, with the largest magnitude of the Ni self-migration barrier, followed by Fe and Cr in a decreasing order, is in line with trends in fcc Fe [47]. Table III also shows that for all species, with the exception of Fe, the $\omega_2$ jumps have considerably lower barriers compared to the corresponding $\omega_{1,3-5}$ barriers. This can be an indication that the species are susceptible to vacancy drag. However, a significantly lower $\omega_2$ barrier can result in a continuous solute-vacancy exchange, with vacancy-trapping causing negligible solute net displacement. For this reason, it is not possible to assess flux-coupling and segregation tendencies of these solutes based solely on the results in Tables I-III. In the case of solute Fe on the other hand, all barriers have similar magnitude, which indicates that the vacancy is less affected by the presence of the species. As it is also shown in Table II that the Fe-vacancy interaction is particularly weak, Fe migration by vacancy drag can be considered less likely compared to other species.

The migration barriers and attempt frequencies of the species considered for interstitial-mediated diffusion are presented in Table IV. A number of migration barriers for Ti and Mn were also included for the sake of consistency.

TABLE IV: Interstitial migration energies, $\omega$ [eV], and attempt frequencies, $\nu$ [THz], for solutes in Ni obtained in this work. The values marked with "†" refer to octahedral jumps, whereas all others regard the dumbbell mechanism. See Fig. 3 for an illustration of the dumbbell jumps and description of notations. All calculations were performed in a 256-atom cell and with an energy cut-off of 350 eV. A migration barrier of 0.0 eV indicates a spontaneous jump without any thermally-activated barrier. Barriers marked with "-" were not calculated in this work.

| | Cr | Fe | P | Si | Ti | Mn |
|---|---|---|---|---|---|---|
| $\omega_0$ Ni SIA jump | 2.04 | | | | | |
| $\omega_0$ Pure Ni jump | 0.14 | | | | | |
| $\omega_{0R}$ Pure Ni rot | 0.61 | | | | | |
| $\omega_1$ Solute jump | 1.77 | 1.42 | 2.21$^\dagger$ | 0.91 | - | - |
| $\omega_1$ Mixed jump | 0.08 | 0.20 | - | 0.03 | 0.004 | 0.13 |
| $\omega_{octa}$ Octahedral-Octahedral | - | - | 0.95$^\dagger$ | - | - | - |
| $\omega_R$ Mixed rot | 0.24 | 0.53 | - | 0.77 | - | - |
| $\omega_{A1}$ 1nnA - 1nnA | 0.09 | 0.17 | - | 0.02 | 0.24 | - |
| $\omega_{A2}$ 1nnA-inf | 0.30 | 0.14 | - | 0.39 | 0.11 | 0.10 |
| $\omega_{A2}'$ inf- 1nnA | 0.14 | 0.17 | - | 0.02 | 0.22 | 0.20 |
| $\omega_{A3}$ 1nnA-mixed | 0.003 | 0.20 | - | 0.042 | 0.33 | 0.12 |
| $\omega_{A3}'$ Mixed -1nnA | 0.29 | 0.15 | - | 0.42 | 0.0 | 0.34 |
| $\omega_{B1}$ 1nn B- 1nnB | 0.17 | 0.15 | - | 0.17 | 0.17 | 0.16 |
| $\omega_{B2}$ 1nnB- inf | 0.23 | 0.18 | - | 0.04 | 0.26 | 0.25 |
| $\omega_{B2}'$ | 0.11 | 0.14 | - | 0.36 | 0.03 | 0.08 |

The migration barrier of 0.95 eV for P octahedral migration is considerably higher than all other barriers in Table IV. This barrier is also significantly higher than that of P octahedra migration in bcc Fe, which is 0.16 eV [51]. One may for this reason not only suspect a far less efficient diffusion of interstitial P in fcc Ni compared to bcc Fe, but also of interstitial P compared to interstitial migration of the other solutes considered in this study - with interstitial Fe being a possible exception. As shown in Table IV, all migration barriers of Fe have similar magnitude, and the values are close to the barrier of the pure-Ni dumbbell jump. Fe is for this reason unlikely to be susceptible to an efficient dumbbell migration. Regarding Cr, the lowest barrier by far is the association jump, $\omega_{A3}$, followed by the mixed-dumbbell jump, $\omega_1$, both of which are lower compared to the pure Ni dumbbell jump, $\omega_0$. In general, the results in Table IV show that the various association jumps of Cr have lower barriers than the respective dissociation jumps. A similar trend is seen for Si, where the only barrier lower than the association jump, $\omega_{A3}$, is the mixed-dumbbell jump, $\omega_1$. Also for this species, the two barriers are considerably lower than those of all other jumps, with the exception of the $\omega_{B2}$ jump. The fact that the Si dissociation barrier is lower for the $\omega_{B2}$ jump can be explained by the repulsive interaction of the species in the 1nnB configuration, as shown in Table II. In Table II it can also be seen that both Cr and Si display strong binding to the ⟨100⟩ dumbbell. The low association barriers, together with the stability of the mixed ⟨100⟩ complexes, are indications of significant dumbbell transport of the two species. However, similarly to the previous discussion for vacancies, a strong correlation can prevent net displacement. Nevertheless, as the mixed-dumbbell migrates via rotation-translation, the change of orientation makes correlation effects weaker as the probability for zero-displacement back jumps is lower compared to that of vacancy migration [31]. Correlation effects are, for this reason, expected to be less important for dumbbell- compared to vacancy-mediated diffusion.

## C. Solute-transport properties

From the DFT data presented in sections 3.1 and 3.2, the transport coefficients of each binary alloy were calculated. In section 3.3.1, the accuracy of the calculated transport coefficients is assessed based on results from experiments. In section 3.3.2, the dominant diffusion mechanisms of the various species during thermal equilibrium are discussed, and in section 3.3.3, radiation-induced segregation tendencies are evaluated.

### *1. Validation with Diffusion Experiments*

As diffusion coefficients can be determined experimentally, these can be used to validate the results of this study. However, is should be noted that experiments are mostly performed in the high-temperature range, whereas the diffusion and transport coefficients computed in this work are based on 0-K DFT parameters. By relying on the 0-K data only, the self-diffusion coefficient, Eq. 4, can be written as:

$$D^* = g a_0^2 f_0 \omega_0 c_V^{eq} \tag{14}$$

where the equilibrium vacancy concentration, $c_V^{eq}$, the vacancy migration rate in pure Ni, $\omega_0$, are given by Eqs 15 and 16.

$$c_V^{eq} = \exp(-E_V^f / k_B T + S_V^f / k_B) \tag{15}$$

$$\omega_0 = \nu_0 \exp(-E_V^{mig} / k_B T) EC_{corr} \tag{16}$$

where $EC_{corr}$ is the correction term taking into account the contribution of electronic excitations, given by Eq. 17.

$$EC_{corr} = \exp(\pi^2 k_B T E_c / 6) \tag{17}$$

where $E_c$ is the difference between the electronic density of states at the saddle point and in the equilibrium on-lattice position [53].

From the 0-K results of this work ($E_V^f = 1.4$ eV, $\nu_0$ = 14.32 THz, and $E_V^{mig} = 1.05$ eV), together with data from Tucker *et al.* ($S_V^f = 1.82$ $k_B$, $E_c$=-0.66 eV) [53], the temperature dependence of the Ni self-diffusion coefficient was calculated. The result is presented in Fig. 5 (red solid line), where it is also compared with experimental self-diffusion coefficients. As can be seen in the figure, calculations compare well with the experimental self-diffusion coefficients at high temperatures, but deviate at lower temperatures. In Table V, the parameters used to obtain the results of Fig. 5 are presented, as well as the resulting activation energies and prefactors. In the table, a considerable mismatch of both activation energy and prefactor can be seen, indicating that the apparent match at high temperatures might be only coincidental. Calculations were also performed using the full set of DFT parameters obtained by Tucker *et al.* ($E_V^f = 1.65$ eV, $E_V^{mig} = 1.09$ eV, $\nu_0$=4.48 THz). The results, represented by the green line in Fig. 5, give a closer match in activation energy, but a more important deviation in prefactor, and thus no significant improvement with respect to the experimental benchmark overall.

TABLE V: Values used to calculate the temperature dependence of the Ni self-diffusion coefficients of Fig. 5.

|  |  | This work | This work+ $C_V^{Gong}$ | Tucker[a] | Fitting of exp data |
|---|---|---|---|---|---|
| Vacancy Formation Energy | $E_V^f$ | 1.40 eV | varying with T[a] | 1.65 eV[a] | - |
| Vacancy Formation Entropy | $S_V^f$ | 1.83 $k_B$[a] | varying with T[b] | 1.83 $k_B$[a] | - |
| Vacancy Migration Energy (eV) | $E_V^{mig}$ | 1.05 | 1.05 | 1.09[a] | - |
| Attempt frequency (THz) | $\nu_0$ | 14.32 | 14.32 | 4.48 THz[a] | - |
| Electronic excitation contribution (eV) | $E_c$ | 0.66[a] | 0.66[a] | 0.66[a] | - |
| Activation energy (eV) (Fig. 5) | Q | 2.44 | 2.49 | 2.74 | 2.83 |
| Prefactor (m$^2$/s) (Fig. 5) | $D_0$ | 7.3×10$^{-6}$ | 7.7×10$^{-6}$ | 2.3×10$^{-6}$ | 5.6×10$^{-5}$ |

[a]Reference [53]
[b]Reference [54]

In the above discussion, a constant vacancy formation free energy was assumed, equal to the value computed at 0 K. However, studies by Gong *et al.* and Glensk *et al.* showed that finite-temperature effects can be important and may lead to strongly non-Arrhenius vacancy concentrations [54,55]. For this reason, an attempt was here made to improve the calculated self-diffusion coefficient by applying Gong's correction. Using the 0-K $\omega_0$ computed in this work, together with the vacancy concentration computed by Gong *et al.*, which accounts for vibrational, electronic, and magnetic finite-temperature contributions to $E_V^f$, the dashed red curve in Fig. 5 was obtained.

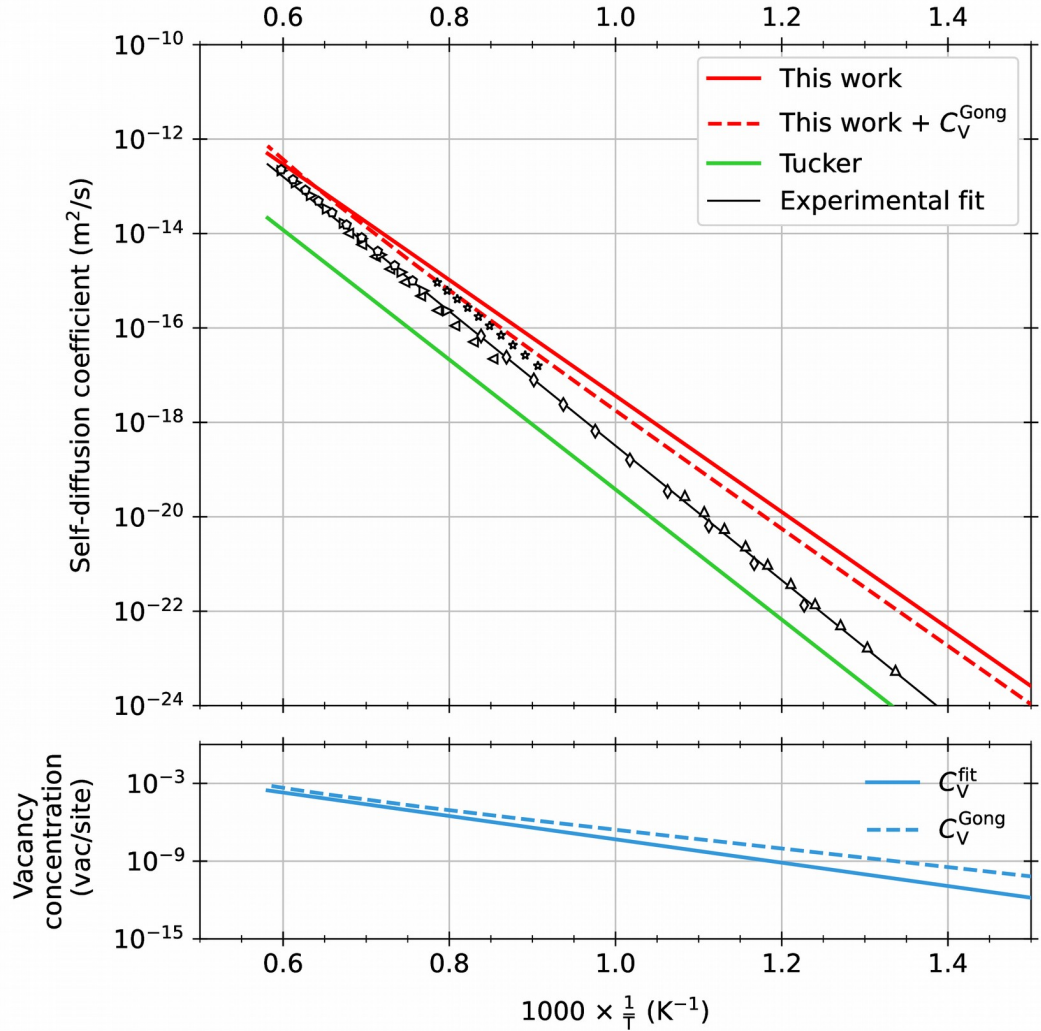

FIG. 5. Ni self-diffusion coefficient ($D^*$) due to thermal vacancies as a function of temperature, compared with experimental data from [58]. The parameters used for each curve correspond to those presented in Table V. The lower inset shows the vacancy concentration necessary to fit the self-diffusion coefficient to the experimental values, when using the 0-K vacancy migration rate $\omega_0$ computed in this work.

The correction with Gong's model improves slightly the match with the experimental diffusion coefficients, both in terms of activation energy and prefactor (Table V); however, a non-negligible deviation remains, especially in the lower-temperature range. The reasons for this mismatch are hard to pinpoint. The finite-temperature effects on the vacancy migration rate, $\omega_0$, neglected in this work are likely to play a role; however, it is also possible that Gong's model, providing a satisfactory match at temperatures close to the melting point (1700 K), does not perform as well at lower temperatures.

It should be noted that calculations in this work are performed with the Generalized Gradient Approximation (GGA). In a study by Hargather *et al.*, the self-diffusion coefficients of fcc Ni was calculated as functions of temperature using the LDA functional [56]. In their work, the authors are able to obtain a very good match with experimentally determined self-diffusion coefficients. The Local Density Approximation (LDA) has been previously shown to predict more accurately the activation energies for diffusion ( [56] and references therein), for which reason this functional has been sometimes used to calculate such parameters. However, as vibrational entropies calculated with the GGA functional are generally considered more reliable [57], this functional is thought to give a more accurate diffusion prefactor. The choice of exchange-correlation functional when calculating diffusion coefficients is for this reason not straight forward. However, diffusion coefficients are proportional to $c_V^{eq}$ (Eq. 5), and as the GGA functional is less accurate in describing this parameter due to the inaccuracy on the vacancy formation entropy, the assumption was here made that the source of error in the calculations of Fig. 5 could be completely ascribed to an incorrect description $c_V^{eq}$. The latter was for this reason treated as a fitting parameter, which was be obtained from the vacancy concentration that fits the DFT-computed values to the experimental diffusion coefficient according to Eq. 18.

$$D_{\exp} = a_0^2 f_0 c_V^{fit} \omega_0 \qquad (18)$$

The fitted vacancy concentration, $c_V^{fit}$, together with the vacancy concentration obtained by the corrections according to the model proposed by Gong *et al.*, are presented in the lower inset of Fig. 5. To verify the accuracy of the assumption that the error in Fig. 5 can be incorporated in $c_V$, the fitted quantity was used to calculate the solute diffusion coefficients of Cr and Fe in dilute Ni, based on the transport coefficients obtained in this work. The results are presented in Fig. 6, together with an Arrhenius fit of experimental values from [58]. An extensive study on solute diffusion in dilute Ni has been previously performed using the 5-frequency model by Hargather *et al.*, [57]. As the authors obtained a very good fit for their calculated Ni self-diffusion coefficient in a previous work [56], their results regarding solute-diffusion of Cr and Fe in Ni are included in the figure for reference. The figure displays a very good match between the results of this work and the experimental values. The current model also significantly improves the estimated solute diffusion coefficients for the two species compared to the Hargather study. These results indicates that the mismatch in Fig. 5 is likely due to an inaccurate estimation of $c_V$ at lower temperatures. For this reason, the solute-related coefficients calculated in this work are considered reliable. In addition, under irradiation, radiation-induced defect formation dominates over the equilibrium defect population, so the role of $c_V^{eq}$ in solute RIS is negligible. Hence, it can be concluded that while the calculation of vacancy equilibrium concentrations in pure Ni is a source of uncertainties, the solute-related properties computed in this work are reliable.

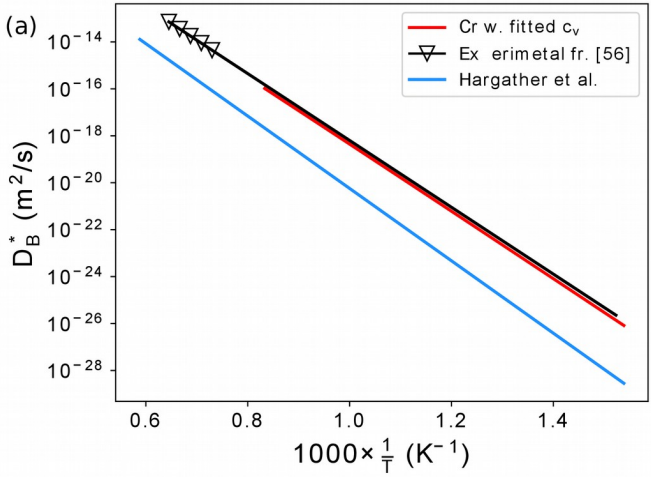

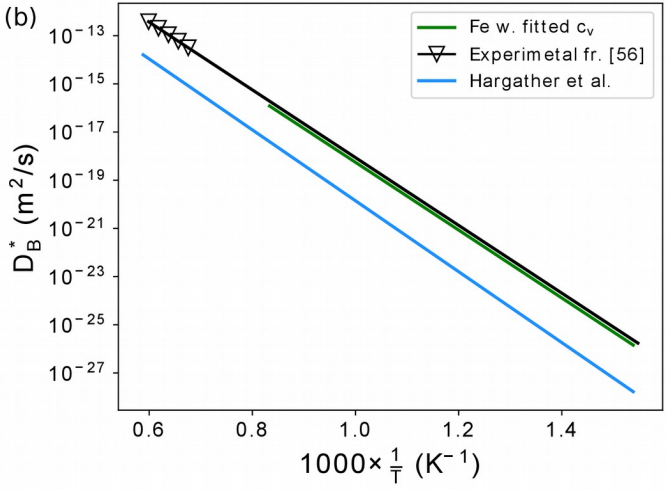

FIG. 6. Vacancy mediated tracer solute diffusion coefficient as a function of temperature for a) Cr b) Fe. Results are compared to experimental data from [58], and with Arrhenius plot obtained with values from Hargather et. al [57]. "fitted $c_V$" has been inferred from the fitting of self-diffusion coefficients shown in Fig. 5.

## *2. Dominant Diffusion Mechanisms*

The dominant diffusion mechanism (vacancy or dumbbell) under thermal conditions can be determined from the ratios of the respective solute tracer diffusion coefficients, $D_{BI}/D_{BV}$. As shown in Eq. 5, the solute diffusion coefficients are proportional to the equilibrium defect concentration. The ratio $c_V/c_I$ can to a first approximation be estimated by $D_I/D_V$ [59], which at low solute concentrations reduces to $L_{II}/L_{VV}$. Using this factor to estimate the ratio of defect concentrations, the preferred defect migration paths for all considered species migrating by both vacancy and interstitial mechanisms were assessed, and results are presented in Fig. 7. In this figure, interstitial P diffusion is based on the octahedral mechanism, which in this case is strongly coupled with the diffusion of pure Ni dumbbells, as the latter are required to kick out P ration is solely based

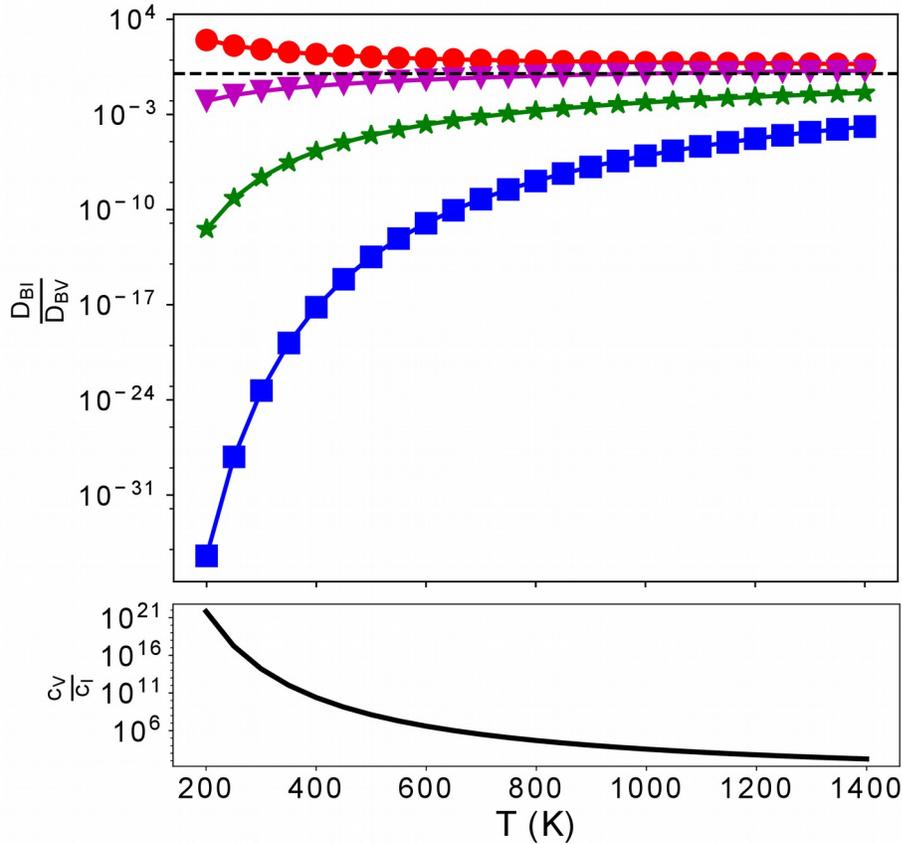

Fig. 7. Ratios of solute tracer diffusion coefficients due to vacancy and interstitial mechanism. As that interstitial diffusion is considered negligible for Ti and Mn, the species are not included in the figure. The lower inset displays the ratio of vacancy- and interstitial concentrations used to calculate the diffusion coefficient ratios.

As can be seen in Fig. 7, vacancy-mediated migration is dominant for all species but Cr, where the dumbbell mechanism is more important. By assuming a constant vacancy-to-solvent fraction of $10^{-6}$, so to simulate irradiation conditions, the solute diffusion coefficients of the preferred mechanisms were computed. As Cr shows a preference to interstitial migration, the interstitial concentration was assessed from the same vacancy ratio, followed by the assumption that $c_V/c_I = L_{II}/L_{VV}$ [59]. Results are presented in Fig. 8. The figure shows that P is by far the fastest diffuser. Interestingly, P has also been shown to be faster than both Cr, Mn, Ni, and Si in bcc Fe [31]. Although in the latter case, the migration of P is dominated by the dumbbell mechanism, fast migration occurs only by vacancies. In Fig. 8 it can be seen that Si is faster than the other solutes with the exception of P. All other solutes have similar diffusion rates to one another, which are also close to that of Ni self-diffusion.

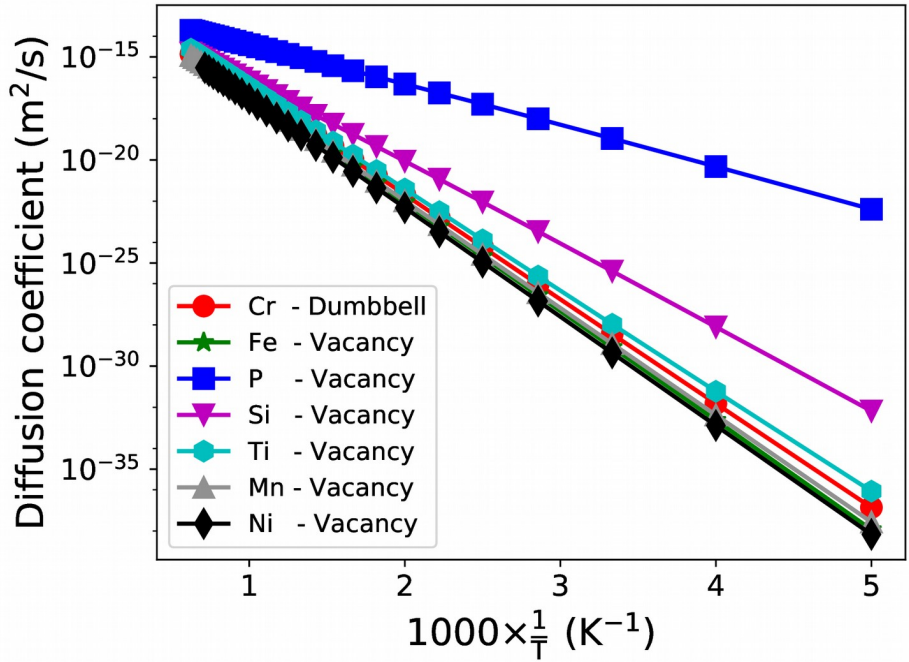

Fig. 8. Diffusion coefficients by the preferred diffusion mechanism for the various species. Calculations were performed while assuming a constant vacancy to solvent fraction of $10^{-6}$.

### 3. Partial Diffusion Coefficients and Radiation-Induced Segregation

In Fig. 9 the conventional vacancy drag ratios, $G_V = L_{VB}^{(VB)}/L_{BB}^{(VB)}$, are presented together with the partial diffusion coefficients (PDCs) for vacancy and dumbbell mediated diffusion. The drag ratio indicates if a solute is likely to follow the vacancy in its migration, $G_V > 1$, or if it will diffuse in the opposite direction, $G_V < 1$. The PDCs show the enrichment/depletion tendency of a solute, induced by one diffusion mechanism while assuming that the other mechanism is inactive. In the case of vacancy-coupled diffusion, enrichment of a species can occur even in the absence of drag, as a result of competition with the surrounding bulk atoms. This is indicated by $0 \leq PDC_{vac} \leq 1$. See section 2.1 for a full description of the PDCs.

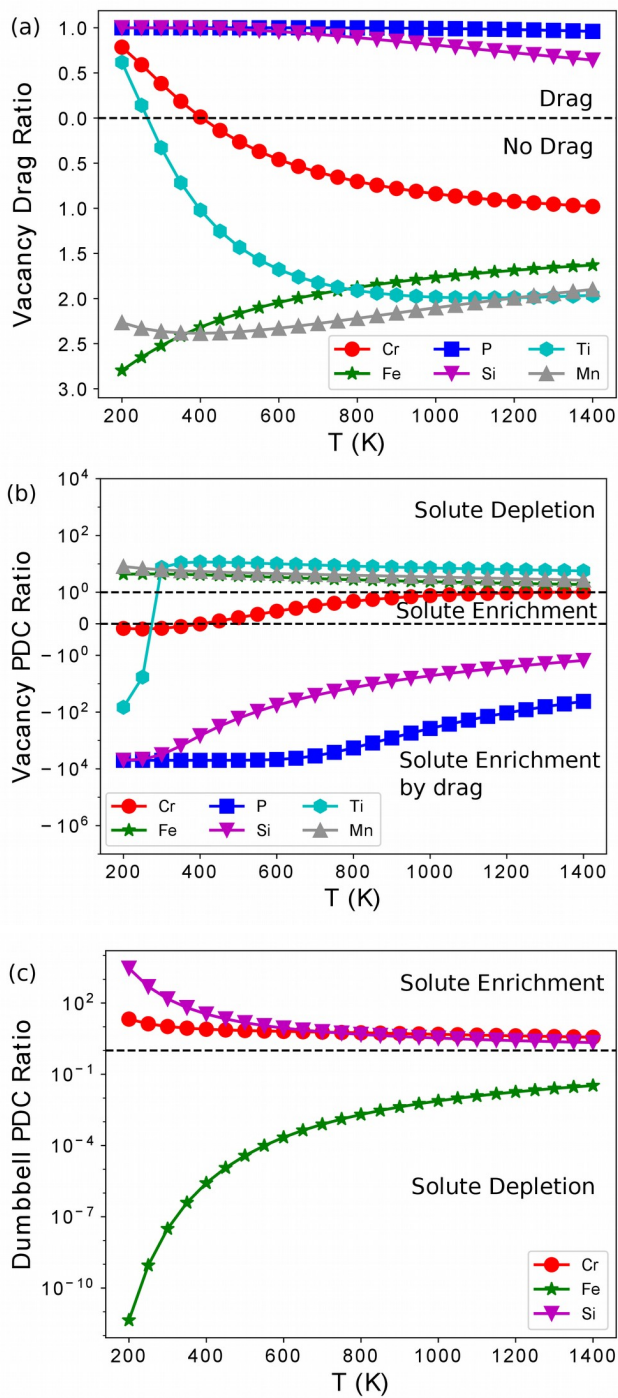

Fig. 9. a) Vacancy drag ratio, $G_V = L_{VB}^{(VB)} / L_{BB}^{(VB)}$, b) Vacancy partial diffusion coefficient ratios, and c) Dumbbell partial diffusion coefficient ratios as functions of temperature. As interstitial diffusion is considered negligible for Ti and Mn, and the octahedral kickout mechanism dominates over the dumbbell mechanism for P, the three are omitted in c).

P and Si are shown in Fig. 9 to enrich at sinks by vacancy drag at all temperatures considered in this study. This result is expected based on observations in sections 3.1 and 3.2, where positive solute-vacancy binding, and a higher probability of vacancy-solute pair association than dissociation, were found for the two species. The combination of the two factors leads to diffusion of the vacancy-solute pairs as coupled species, thus explaining the vacancy-mediated enrichment. In the case of Ti, results in Fig. 9 show a switchover at 320 K between enrichment due to vacancy drag, and depletion in the absence of drag. Table II indicates a weak 1nn binding interaction between Ti and vacancies, and in Table III it is shown that the Ti-vacancy migration barrier, $\omega_2$, is approximately 50% as high as the $\omega_{1,3\text{-}5}$ barriers. As discussed in section 3.2, if $\omega_2$ is considerably lower than the other barriers, preferential solute-vacancy exchange can prevent net transport in the material. This behaviour is incorporated in the correlation factor, $f_B$ of Eq. 5, whose temperature dependence follows an Arrhenius behaviour and can, for this reason, be seen as an additional contribution to the migration activation energy. For Ti, this contribution was calculated to $E_f$=0.52 eV, which in combination with the original $\omega_2$ results in an effective barrier of 1.02 eV. This can be expected to slow down diffusion, but as the vacancy diffusion rate is affected to an equal extent as that of the solute, the effective barrier does not explain the crossover in drag/no drag demonstrated in Fig. 9. This behaviour is instead likely due to the combination of a weak Ti-1nn binding and Ti-2nn repulsion. Indeed, a similar switchover has been observed for vacancy mediated diffusion of Ti in bcc Fe, where a weak 1nn binding and a 2nn repulsion have also been demonstrated [60].

In Fig. 9, it can be seen that both Fe and Mn are depleted due to the inverse-Kirkendall mechanism, in which vacancies and solutes move in opposite directions. The opposite behaviour is observed in the lower temperature range for Cr, which is dragged by vacancies up to a crossover temperature of 400 K. However, as indicated by Fig. 9(b), Cr is enriched up to temperatures of approximately 1000 K as a consequence of Ni being more effective at diffusing away from the sinks. The above results are in line with previous results by Tucker *et al.*, where a crossover temperature of approximately 460 K was obtained for Cr drag by vacancies, and the inverse-Kirkendall mechanism was found for Fe at all temperatures [53].

In the case of dumbbell mediated diffusion, Fig. 9(c) displays enrichment of Cr and Si, and depletion of Fe. As can be seen in Table II, both Si and Cr form stable mixed ⟨100⟩ dumbbells. Table IV also shows that Cr and Si generally have lower association energy barriers compared to those of dissociation. Fe, on the other hand repels the dumbbell configuration, whereas the various association/dissociation barriers are similar for this species. Thus results seen in Fig. 9(c) are in line with what could have been expected based on the DFT study in the first part of this work. It should be noted that the Fe PDC$_{dumb}$ is very small compared to those of Si and Cr. This result supports the discussion of section 3.1, according to which repulsive interactions of the mixed Ti and Mn ⟨100⟩ dumbbells were seen as indication of negligible SIA diffusion for the two species. Since the SIA-Ti and the SIA-Mn interactions are even more negative compared to Fe, the two will confidently not be subjected to interstitial diffusion.

Once the segregation tendencies of vacancy- and interstitial mediated diffusion have been determined through the PDCs, the overall RIS behaviour can be estimated. RIS tendencies are described by α, defined in Eq. 3. If the factor is positive, enrichment of the solute occurs at defect sinks, whereas a negative factor describes depletion. The RIS factors of the solutes considered in this work are presented in Fig. 10 as functions of temperature. Ti and Mn were assumed not to have any transport by interstitials, which was accounted for by setting the corresponding $L_{IB}$ and $L_{BB}$ of Eq. 3 to zero, and $L_{II}$ is that of pure Ni dumbbell migration. This makes the RIS factors for the two species independent of interstitial migration. In the case of P interstitial migration, only the octahedral configuration was considered. As discussed in section 2.2, this configuration forms spontaneously if a migrating Ni dumbbell encounters a substitutional P in the lattice, as one of the Ni atoms takes the original place of the P atom.

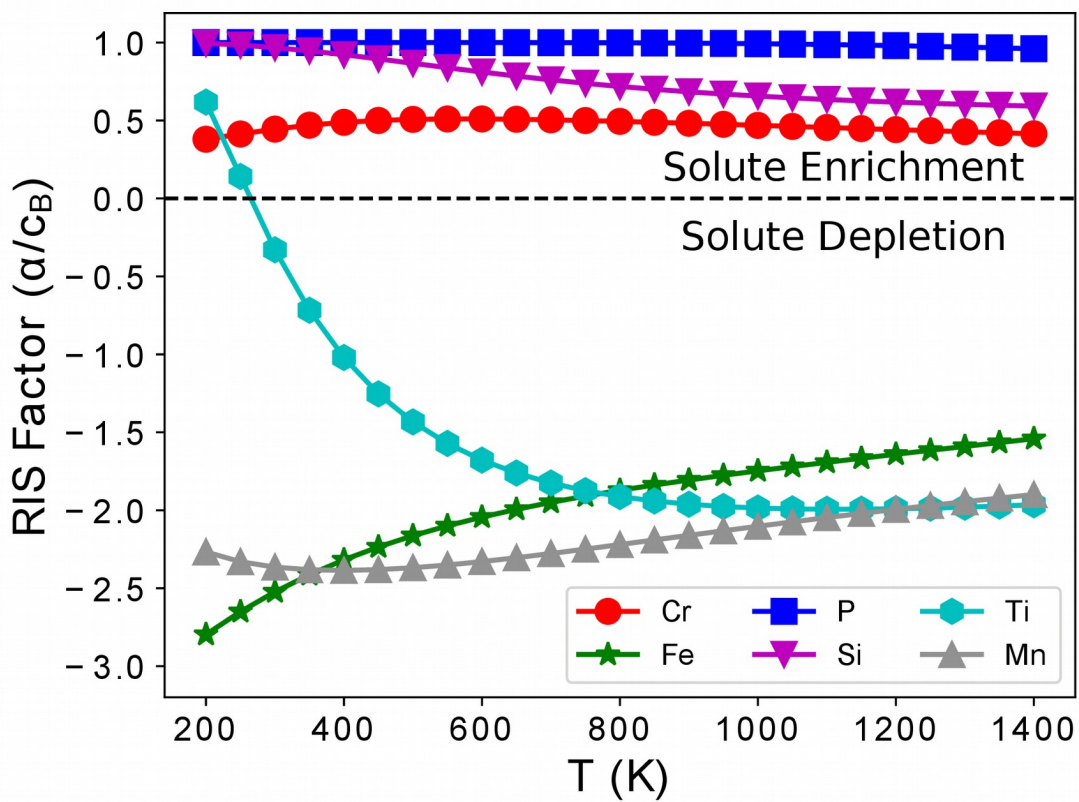

Fig. 10. RIS tendencies of Cr, Fe, P, Si, Ti, and Mn in bulk nickel. Positive values indicate enrichment at sinks.

# IV. DISCUSSION

Previous studies of austenitic alloys show Ni, Si, and P enrichment, and Fe, Cr, Mn depletion at grain boundaries and interfaces due to RIS [61–69]. In Ni alloys, Cr enrichment and Fe depletion have been detected at sinks [67,70,71], which is in line with the results of this work. As shown in Fig. 10, Fe and Mn are the only species which are depleted at all temperatures, since both diffuse via the inverse-Kirkendall mechanism. Interstitial diffusion is shown in Fig 9 to give rise to Fe depletion, resulting from the influx of pure Ni dumbbells towards sinks. Based on the magnitudes of the PDC ratios in Fig 9, it can however be assumed that the interstitial contribution is less important than that of the vacancy mechanism for the overall RIS of Fe in fcc Ni (Eq. 3). As the results in Fig. 7 show that the vacancy mechanism is also dominant in the absence of radiation, it can be concluded that in the case of Fe migration in Ni, the vacancy mechanism is dominant both in thermal and non-equilibrium conditions. Results regarding Fe in this study are not directly applicable to fcc Fe alloys, as the behaviour of Fe in Ni can be very different from the behaviour of Ni in Fe. However, in the case of Ni precipitation in austenitic steels, Fe atoms inside such a precipitate will see a Ni-rich environment. In this case, Fe is likely to behave as predicted in this study, i.e., it will deplete from the precipitate surface (which is a sink for defects), and agglomerate towards the centre of the precipitate.

In the case of Si, results in this work show that both interstitial and vacancy mechanisms contribute to its enrichment. Enrichment at sinks has been observed experimentally following irradiation of Ni and fcc Fe alloys [15,16,62,63,72,73]. From the PDC ratios of this work, one can conclude that the vacancy mechanism dominates the observed behaviour up to temperatures of ~1200K, where the interstitial mechanism becomes slightly more important. A similar behaviour is observed in thermal conditions, as Fig. 7 displays a slight preference for the vacancy mechanism, although the difference is negligible at higher temperatures. One can conclude that the experimental observations are likely due to the vacancy mechanism. As the same mechanism dominates Si migration in irradiated bcc Fe [31], one can suspect a pronounced vacancy contribution to the observed Si enrichment in fcc Fe alloys.

P displays enrichment due to both vacancy and octahedral mechanisms. From results in Fig. 7, and the magnitudes of the PDC ratios in Fig. 9, it can be concluded that P diffusion is dominated by the vacancy mechanism. This is consistent with the significant migration barrier, 0.95 eV, for octahedral diffusion. Segregation of P has also been observed in bcc Fe, and is a well-known problem in austenitic steels. The preferred migration mechanism of P is different in fcc Ni compared to bcc Fe, where enrichment of P is dominated by the dumbbell mechanism [31]. However, P displays strong tendencies for vacancy drag in both materials. As the mixed Ni-P dumbbell is not stable in fcc Ni, one can suspect that the vacancy mechanism is more important in austenitic steel compared to bcc Fe. However, the stability of the various mixed P-X dumbbells in austenitic alloys remains to be verified in order to safely determine the relative importance of the two mechanisms in the material.

Based on DFT results in section 3.1, it can be concluded that Ti does not migrate via the interstitial mechanism, and that only the vacancy mechanism contributes to RIS for this species. In bcc Fe, Ti diffusion by dumbbells has been shown to be negligible due to repulsion of the Fe-Ti dumbbell [60]. Additionally, the crossover for enrichment/depletion at 320 K displayed in Fig. 10, resembles observations of Ti in bcc Fe, where a switchover between enrichment and depletion has been found at approximately 700 K [60]. Thus, the behaviour of Ti in fcc Ni and in bcc Fe bears significant similarities. However, the crossover temperature of 320 K in Ni is well below reactor operating temperatures, whereas the crossover temperature in Fe is not. This gives rise to difficulties in transposing observations to austenitic alloys used in current NPPs. Nevertheless, Ti has previously been shown to prevent swelling of austenitic materials in reactor applications [74–77] and to have a stabilizing effect on voids in Ni-based model alloys [78]. Results in this work indicate that a possible explanation for these observations is the trapping of vacancies due to the preferential exchange with Ti. A more thorough investigation of vacancy trapping by Ti would entail the computation of the vacancy diffusion coefficient as a function of Ti concentration, but this is beyond the scope of this work.

In Fig. 10, Cr is shown to enrich at sinks as a consequence of both to the interstitial and vacancy mechanisms. Based on diffusion coefficients and PDC ratios, it can be concluded that Cr preferentially diffuses by the dumbbell migration. Previous studies of Cr in fcc Ni generally detect depletion at sinks, however enrichment has been observed in a work by Allen et al. [65]. In their study, RIS in three different Ni based alloys (Ni-18Cr, Ni-18Cr-9Fe, and Ni-18Cr-0.08P) was examined by exposing the materials to either thermal treatment or proton irradiation. Cr depletion was observed for all samples, with the exception of one, where enrichment at sinks was observed. The authors of the study suggested that the behaviour of Cr is very sensitive to interactions with the surroundings. In bcc Fe, the dumbbell mechanism results in enrichment and vacancy mechanism in depletion of Cr near sinks [31,79]. However, the dominant mechanism and overall evolution of Cr were again reported to be very sensitive to temperature, sink density, Cr concentration, and local strain fields [31,79,80]. The sensitivity of Cr to external conditions, both in fcc Ni and bcc Fe, gives rise to difficulties in transposing this work to fcc Fe and its alloys.

Results here presented indicate that vacancy-mediated diffusion is dominant for all species in fcc Ni, with the exception of Cr where the dumbbell mechanism is prominent. The dominant diffusion mechanism does not change for either species depending on whether thermal or non-equilibrium conditions are considered. Interestingly, the vacancy dominated migration in Ni is in line with the common explanation that RIS in austenitic Fe alloys occurs as a consequence of the inverse Kirkendall mechanism [81–83].

Characterizations of irradiated austenitic alloys following reactor service show a high number density of Ni-Si enriched and Cr-Fe depleted clusters, together with P segregation at the interfaces of the clusters [62] or RIS-induced Ni enrichment in the vicinity of grain boundaries. Although not directly applicable to austenitic alloys, results in the current study are relevant for transport phenomena occurring inside and near the internal surface of such clusters or in the vicinity of segregated grain boundaries. A natural continuation of this work would be to perform a similar study in fcc Fe, as this would give a more precise insight on the impact of the lattice structure on segregation tendencies. In future works, the effect of local composition could also be included in order to evaluate the segregation tendencies as functions of solute concentration. With that said, this study provides valuable information for improving the current understanding of RIS in Ni-based alloys.

# V. CONCLUSION

Solute diffusion, flux coupling, and radiation induced segregation (RIS) of Fe, Cr, P, Si, Ti, and Mn in fcc Ni have here been investigated by coupling first-principles calculations with the self-consistent mean field theory. The goal has been to improve the current understanding of radiation-induced segregation processes of materials commonly used in today's and future generation nuclear power plants. For this reason, findings have been compared with similar studies in bcc Fe, and discussed in the context of RIS in austenitic steel. Results show that interstitial migration has little impact on solute diffusion in fcc Ni compared to that of vacancies, with the exception of Cr, for which the migration is dominated by a dumbbell mechanism. In addition, it has here been shown that P, Si and Cr are enriched at sinks as a consequence of radiation-induced segregation, whereas Fe and Mn are depleted. Ti was shown to enrich at sinks at low temperatures, with a switchover near room temperature, followed by depletion in the higher temperature range. Results in fcc Ni are to a great extent in line with observations in bcc Fe, where Si and Ti migration are dominated by the vacancy mechanism, and Cr dumbbell migration leads to enrichment at sinks. Interstitial P however, behaves very differently in fcc Ni compared to bcc Fe. In Fe, the solute forms a quickly-migrating, hardly dissociating mixed dumbbell, whereas in Ni, substitutional P will preferentially be kicked out into a more slowly-migrating octahedral configuration.

This work has improved the understanding of the underlying atomic-transport phenomena behind solute segregation in irradiated materials. As calculations were performed in the dilute limit, the results are not directly applicable to concentrated Ni alloys in current nuclear power plants. However, this is the first comprehensive modelling-based analysis of RIS tendencies due to both vacancy and dumbbell mechanisms in Ni alloys. The results shed a light on the intrinsic kinetic behaviour of several important solutes in Ni alloys and austenitic steels, and can for this reason be considered an important milestone towards a broader picture of irradiation damage in nuclear structural materials.

# ACKNOWLEDGEMETS

This work is currently financed by the Euratom research and training programme 2014-2018 under grant agreement No 755269. We acknowledge EDF R&D HPC resources as well as the CINECA award under the ISCRA initiative, for the availability of high performance computing resources and support. The computations were also enabled by resources provided by the Swedish National Infrastructure for Computing (SNIC) at KTH Royal Institute of Technology partially funded by the Swedish Research Council through grant agreement no. 2016-07213

# BIBLIOGRAPHY

[1]  D. Lemarchand, E. Cadel, S. Chambreland, and D. Blavette. Philos. Mag. A **82**, 9, (2002).
[2]  A. F. Rowcliffe, L. K. Mansur, D. T. Hoelzer, and R. K. Nanstad. J. Nucl. Mater. **392**, 2, (2009).

[3] J. L. Brimhall, D. R. Baer, and R. H. Jones. J. Nucl. Mater. **117**, (1983).
[4] K. Jin, W. Guo, C. Lu, M. W. Ullah, Y. Zhang, W. J. Weber, L. Wang, J. D. Poplawsky, and H. Bei. Acta Mater. **121**, (2016).
[5] M. W. Ullah, D. S. Aidhy, Y. Zhang, and W. J. Weber. Acta Mater. **109**, (2016).
[6] D. Chakraborty and D. S. Aidhy. J. Alloys Compd. **725**, (2017).
[7] Y. Yang, K. G. Field, T. R. Allen, and J. T. Busby. J. Nucl. Mater. **473**, (2016).
[8] E. Levo, F. Granberg, C. Fridlund, K. Nordlund, and F. Djurabekova. J. Nucl. Mater. **490**, (2017).
[9] T. Yang, C. Lu, G. Velisa, K. Jin, P. Xiu, M. L. Crespillo, Y. Zhang, H. Bei, and L. Wang. Acta Mater. **151**, (2018).
[10] C. Lu, K. Jin, L. K. Béland, F. Zhang, T. Yang, L. Qiao, Y. Zhang, H. Bei, H. M. Christen, R. E. Stoller, and L. Wang. **6**, (2016).
[11] K. Sato, D. Itoh, T. Yoshiie, Q. Xu, A. Taniguchi, and T. Toyama. J. Nucl. Mater. **417**, 1, (2011).
[12] T. Yoshiie, Q. Xu, Y. Satoh, H. Ohkubo, and M. Kiritani. J. Nucl. Mater. **283–287**, (2000).
[13] T. Yoshiie, T. Ishizaki, Q. Xu, Y. Satoh, and M. Kiritani. J. Nucl. Mater. **307–311**, (2002).
[14] K. Hamada, S. Kojima, Y. Ogasawara, T. Yoshiie, and M. Kiritani. J. Nucl. Mater. **212–215**, (1994).
[15] L. E. Rehn, P. R. Okamoto, and H. Wiedersich. J. Nucl. Mater. **80**, 1, (1979).
[16] R. S. Averback, L. E. Rehn, W. Wagner, H. Wiedersich, and P. R. Okamoto. Phys. Rev. B **28**, 6, (1983).
[17] M. Yamaguchi, M. Shiga, and H. Kaburaki. J. Phys. Condens. Matter **16**, 23, (2004).
[18] M. Všianská and M. Šob. Prog. Mater. Sci. **56**, 6, (2011).
[19] S. G. Druce, G. Gage, and G. Jordan. Acta Metall. **34**, 4, (1986).
[20] W. T. Geng, A. J. Freeman, R. Wu, C. B. Geller, and J. E. Raynolds. Phys. Rev. B **60**, 10, (1999).
[21] J. Dong, M. Zhang, X. Xie, and R. G. Thompson. Mater. Sci. Eng. A **328**, 1, (2002).
[22] L. E. Rehn, P. R. Okamoto, D. I. Potter, and H. Wiedersich. J. Nucl. Mater. **74**, 2, (1978).
[23] E. T. Bentria, I. K. Lefkaier, A. Benghia, B. Bentria, M. B. Kanoun, and S. Goumri-Said. Sci. Rep. **9**, 1, (2019).
[24] Packan, N., R. Stoller, A. Kumar, *Effects of Radiation on Materials: 14th International Symposium (Volume I)* (ASTM International 1990).
[25] A. R. Allnatt and A. B. Lidiard, *Atomic Transport in Solids* (Cambridge University Press, 2003).
[26] M. Nastar. Philos. Mag. **85**, 32, (2005).
[27] T. Garnier, D. R. Trinkle, M. Nastar, and P. Bellon. Phys. Rev. B **89**, 14, (2014).
[28] D. R. Trinkle. Philos. Mag. **97**, 28, (2017).
[29] V. Barbe and M. Nastar. Philos. Mag. **86**, 23, (2006).
[30] T. Schuler, L. Messina, and M. Nastar. Comput. Mater. Sci. **172**, (2020).
[31] L. Messina, T. Schuler, M. Nastar, M.-C. Marinica, and P. Olsson. Acta Mater. **191**, (2020).
[32] A. D. Le Claire. J. Nucl. Mater. **69–70**, (1978).
[33] M. Nastar and F. Soisson, *Comprehensive Nuclear Materials* (Elsevier: Oxford, 2012), Vol. 2, p. 471–496.
[34] E. Martínez, O. Senninger, A. Caro, F. Soisson, M. Nastar, and B. P. Uberuaga. Phys. Rev.

Lett. **120**, 10, (2018).

[35] T. Schuler and M. Nastar. Phys. Rev. B **93**, 22, (2016).
[36] A. Van der Ven and G. Ceder. Phys. Rev. Lett. **94**, 4, (2005).
[37] G. Kresse and J. Hafner. Phys. Rev. B **47**, 1, (1993).
[38] G. Kresse and J. Furthmüller. Phys. Rev. B **54**, 16, (1996).
[39] J. P. Perdew, K. Burke, and M. Ernzerhof. Phys. Rev. Lett. **77**, 18, (1996).
[40] G. Mills, H. Jónsson, and G. K. Schenter. Surf. Sci. **324**, 2, (1995).
[41] G. H. Vineyard. J. Phys. Chem. Solids **3**, 1, (1957).
[42] G. Henkelman, B. P. Uberuaga, and H. Jónsson. J. Chem. Phys. **113**, 22, (2000).
[43] A. Togo and I. Tanaka. Scr. Mater. **108**, (2015).
[44] R. Nazarov, T. Hickel, and J. Neugebauer. Phys. Rev. B **85**, 14, (2012).
[45] R. A. Johnson and A. N. Orlov. *Physics of Radiation Effects in Crystals* (Elsevier, 2012).
[46] C.-C. Fu, F. Willaime, and P. Ordejón. Phys. Rev. Lett. **92**, 17, (2004).
[47] T. P. C. Klaver, D. J. Hepburn, and G. J. Ackland. Phys. Rev. B **85**, 17, (2012).
[48] J. B. Piochaud, T. P. C. Klaver, G. Adjanor, P. Olsson, C. Domain, and C. S. Becquart. Phys. Rev. B **89**, 2, (2014).
[49] C. Domain and C. S. Becquart. Phys. Rev. B **71**, 21, (2005).
[50] F. Maury, P. Lucasson, A. Lucasson, F. Faudot, and J. Bigot. **17**, 5, (1987).
[51] E. Meslin, C.-C. Fu, A. Barbu, F. Gao, and F. Willaime. Phys. Rev. B **75**, 9, (2007).
[52] W. Meyer and H. Nelder. Z Tech 18 (1937).
[53] J. D. Tucker, R. Najafabadi, T. R. Allen, and D. Morgan. J. Nucl. Mater. **405**, 3, (2010).
[54] Y. Gong, B. Grabowski, A. Glensk, F. Körmann, J. Neugebauer, and R. C. Reed. Phys. Rev. B **97**, 21, (2018).
[55] A. Glensk, B. Grabowski, T. Hickel, and J. Neugebauer. Phys. Rev. X **4**, 1, (2014).
[56] C. Z. Hargather, S.-L. Shang, Z.-K. Liu, and Y. Du. Comput. Mater. Sci. **86**, (2014).
[57] C. Z. Hargather, S.-L. Shang, and Z.-K. Liu. Acta Mater. **157**, (2018).
[58] H. Mehrer, *Diffusion in Solid Metals and Alloys* (Springer-Verlag, Berlin/Heidelberg, 1990), Vol. 26.
[59] R. Sizmann. J. Nucl. Mater. **69–70**, (1978).
[60] L. Messina, M. Nastar, N. Sandberg, and P. Olsson. Phys. Rev. B **93**, 18, (2016).
[61] G. S. Was, J. P. Wharry, B. Frisbie, B. D. Wirth, D. Morgan, J. D. Tucker, and T. R. Allen. J. Nucl. Mater. **411**, 1, (2011).
[62] A. Etienne, B. Radiguet, P. Pareige, J.-P. Massoud, and C. Pokor. J. Nucl. Mater. **382**, 1, (2008).
[63] A. Etienne, B. Radiguet, N. J. Cunningham, G. R. Odette, and P. Pareige. J. Nucl. Mater. **406**, 2, (2010).
[64] C. Lu, T. Yang, K. Jin, N. Gao, P. Xiu, Y. Zhang, F. Gao, H. Bei, W. J. Weber, K. Sun, Y. Dong, and L. Wang. Acta Mater. **127**, (2017).
[65] T. R. Allen, L. Tan, G. S. Was, and E. A. Kenik. J. Nucl. Mater. **361**, 2, (2007).
[66] N. Sakaguchi, H. Takahashi, and H. Ichinose. Mater. Trans. **46**, 3, (2005).
[67] L. Barnard, J. D. Tucker, S. Choudhury, T. R. Allen, and D. Morgan. J. Nucl. Mater. **425**, 1, (2012).
[68] M. Hatakeyama, S. Tamura, and I. Yamagata. Mater. Lett. **122**, (2014).
[69] Z. Jiao and G. S. Was. Acta Mater. **59**, 3, (2011).

[70] X.-X. Wang, L.-L. Niu, and S. Wang. Mater. Lett. **202**, (2017).
[71] A. Barashev, Y. Osetsky, H. Bei, C. Lu, L. Wang, and Y. Zhang. Curr. Opin. Solid State Mater. Sci. **23**, 2, (2019).
[72] T. Ezawa, E. Wakai, and R. Oshima. J. Nucl. Mater. **283–287**, (2000).
[73] P. K. Rastogi and A. J. Ardell. Acta Metall. **19**, 4, (1971).
[74] D. J. Mazey, D. R. Harries, and J. A. Hudson. J. Nucl. Mater. **89**, 1, (1980).
[75] A. Hishinuma and K. Fukai. J. Nucl. Sci. Technol. **20**, 8, (1983).
[76] R. M. Boothby and T. M. Williams. J. Nucl. Mater. **152**, 2, (1988).
[77] H. Kawanishi, M. Yamada, K. Fukuya, and S. Ishino. J. Nucl. Mater. **104**, (1981).
[78] K. Ma, B. Décamps, A. Fraczkiewicz, F. Prima, and M. Loyer-Prost. Mater. Res. Lett. **8**, 5, (2020).
[79] O. Senninger, F. Soisson, E. Martínez, M. Nastar, C.-C. Fu, and Y. Bréchet. Acta Mater. **103**, (2016).
[80] F. Soisson, E. Meslin, and O. Tissot. J. Nucl. Mater. **508**, (2018).
[81] T. R. Allen, J. T. Busby, G. S. Was, and E. A. Kenik. J. Nucl. Mater. **255**, 1, (1998).
[82] T. R. Allen and G. S. Was. Acta Mater. **46**, 10, (1998).
[83] J. P. Wharry and G. S. Was. J. Nucl. Mater. **442**, 1, (2013).